\documentclass[
superscriptaddress,
amsmath,
amssymb,
aps,
prl,
twocolumn,
floatfix
]{revtex4-1}

\def\kr{k_{\rm R}}                                    
\def\Er{E_{\rm R}}                                    
\def\Rb87{^{87}\mathrm{Rb}}                     
\def\K40{^{40}\mathrm{K}}                       

\def\ex{\mathbf{e}_x}

\def\ez{\mathbf{e}_z}

\def\bra#1{\mathinner{\langle{#1}|}}
\def\ket#1{\mathinner{|{#1}\rangle}}

{\catcode`\|=\active
  \gdef\Braket#1{\left<\mathcode`\|"8000\let|\BraVert {#1}\right>}}
\def\BraVert{\egroup\,\mid@vertical\,\bgroup}

\def\shorttimes{\!\times\!}

\usepackage{graphicx}          
\usepackage{dcolumn}           
\usepackage{bm}                
\usepackage{units}             
\usepackage[usenames,dvipsnames]{color}
\usepackage[colorlinks,urlcolor=blue,citecolor=blue,linkcolor=blue]{hyperref}
\usepackage{upgreek}
\usepackage[T1]{fontenc}
\usepackage[utf8]{inputenc}

\begin{document}

\title{Realization of a fractional period adiabatic superlattice}

\author{R. P. Anderson}
\thanks{These two authors contributed equally.}
\affiliation{Joint Quantum Institute, University of Maryland, College Park, Maryland, 20742, USA}
\affiliation{School of Physics and Astronomy, Monash University, Melbourne, Victoria 3800, Australia}
\affiliation{La Trobe Institute of Molecular Science, La Trobe University, Bendigo, Victoria 3552, Australia}

\author{D. Trypogeorgos}
\thanks{These two authors contributed equally.}
\affiliation{INO-CNR BEC Center and Dipartimento di Fisica, Universit\`a di Trento, 38123 Povo, Italy}
\affiliation{Joint Quantum Institute, University of Maryland, College Park, Maryland, 20742, USA}

\author{A. Vald\'es-Curiel}
\author{Q.-Y. Liang}
\author{J. Tao} 
\author{M. Zhao}
\affiliation{Joint Quantum Institute, University of Maryland, College Park, Maryland, 20742, USA}

\author{T. Andrijauskas}
\author{G. Juzeli\={u}nas}
\affiliation{Institute of Theoretical Physics and Astronomy, Vilnius University, Saul\.{e}tekio 3, LT-10257 Vilnius, Lithuania}

\author{I. B. Spielman}
\email{ian.spielman@nist.gov}
\homepage{http:// ultracold.jqi.umd.edu}
\affiliation{Joint Quantum Institute, University of Maryland, College Park, Maryland, 20742, USA}
\affiliation{National Institute of Standards and Technology, Gaithersburg, Maryland 20899, USA}

\date{\today}%

\begin{abstract}
We propose and realize a deeply sub-wavelength optical lattice for ultracold neutral atoms using $N$ resonantly Raman-coupled internal degrees of freedom.
Although counter-propagating lasers with wavelength $\lambda$ provided two-photon Raman coupling, the resultant lattice-period was $\lambda/2N$, an $N$-fold reduction as compared to the conventional $\lambda/2$ lattice period.
We experimentally demonstrated this lattice built from the three $F=1$ Zeeman states of a $\Rb87$ Bose-Einstein condensate, and generated a lattice with a $\lambda/6=\unit[132]{nm}$ period from $\lambda=\unit[790]{nm}$ lasers.  Lastly, we show that adding an additional RF coupling field converts this lattice into a superlattice with $N$ wells uniformly spaced within the  original $\lambda/2$ unit cell. 
\end{abstract}

\maketitle

%
%

Optical lattices form a vital substrate for quantum-gases, enabling: the quantum simulation of iconic condensed matter systems~\cite{Greiner2002}, the realization of new atomic topological materials~\cite{Wu2016}, and new-generation atomic clocks~\cite{Bloom2014}.
Generally the spatial period of an optical lattice is derived from the difference of the wave-vectors of the underlying laser beams (with wavelength $\lambda$), forging an apparent lower limit of $\lambda/2$ to the lattice-period.
Many techniques can add sub-wavelength structure to a lattice, ranging from Raman methods~\cite{gupta_raman-induced_1996}, radio-frequency-dressed state-dependent optical lattices~\cite{Lundblad2008, Yi2008}, and time-modulated ``Floquet'' lattices~\cite{Nascimbene2015}, to deeply sub-wavelength structures using dark states~\cite{Lacki2016,Jendrzejewski2016,wang_dark_2018}.
Going beyond these techniques, only explicit use of multi-photon transitions has to-date reduced the underlying lattice period in quantum-gas experiments~\cite{weitz_optical_2004, Zhang2005, Ritt2006}.
In the broader setting, sub-wavelength optical structures may defeat the diffraction limit for lithography, using either non-classical light~\cite{Haske2007}, or even coherent atomic dynamics~\cite{You2018}.
Here, we propose and demonstrate a flexible sub-wavelength lattice with period $\lambda/2N$ built from $N$ resonantly coupled atomic states; furthermore, an additional coupling field converts the lattice to a tunable $N$-well superlattice with $\lambda/2$ periodicity.

Any 1D lattice can be described by a Hamiltonian $\hat H(\hat x) = \hat H(\hat x + \delta x)$ that is invariant under spatial displacements $\delta x$.
The smallest such displacement $d$ defines the lattice's unit cell, and correspondingly $\hat H(\hat x)$ couples momentum states differing by integer multiples of the resulting reciprocal lattice constant $k_0 = 2\pi/d$.
An optical lattice with spatial period $\lambda/2$ formed by a pair of counter-propagating lasers with wavelength $\lambda$ and single-photon momentum $\hbar \kr = 2\pi\hbar /\lambda$ is intuitively derived from the $\hbar k_0 = 2 \hbar \kr$ momentum obtained by exchanging photons between lattice-lasers.
This 2-photon concept has been directly extended to higher order 4- or even 6-photon transitions producing lattices with reduced period~\cite{weitz_optical_2004, Zhang2005, Ritt2006}, but required a concomitant increase in laser intensities.
In this work, we exploit a hidden gauge symmetry present for $N$ internal atomic states coupled by conventional two-photon Raman transitions to generate a highly tunable lattice with period $\lambda/2N$.

%
%

\begin{figure}[t]
\begin{center}
\includegraphics{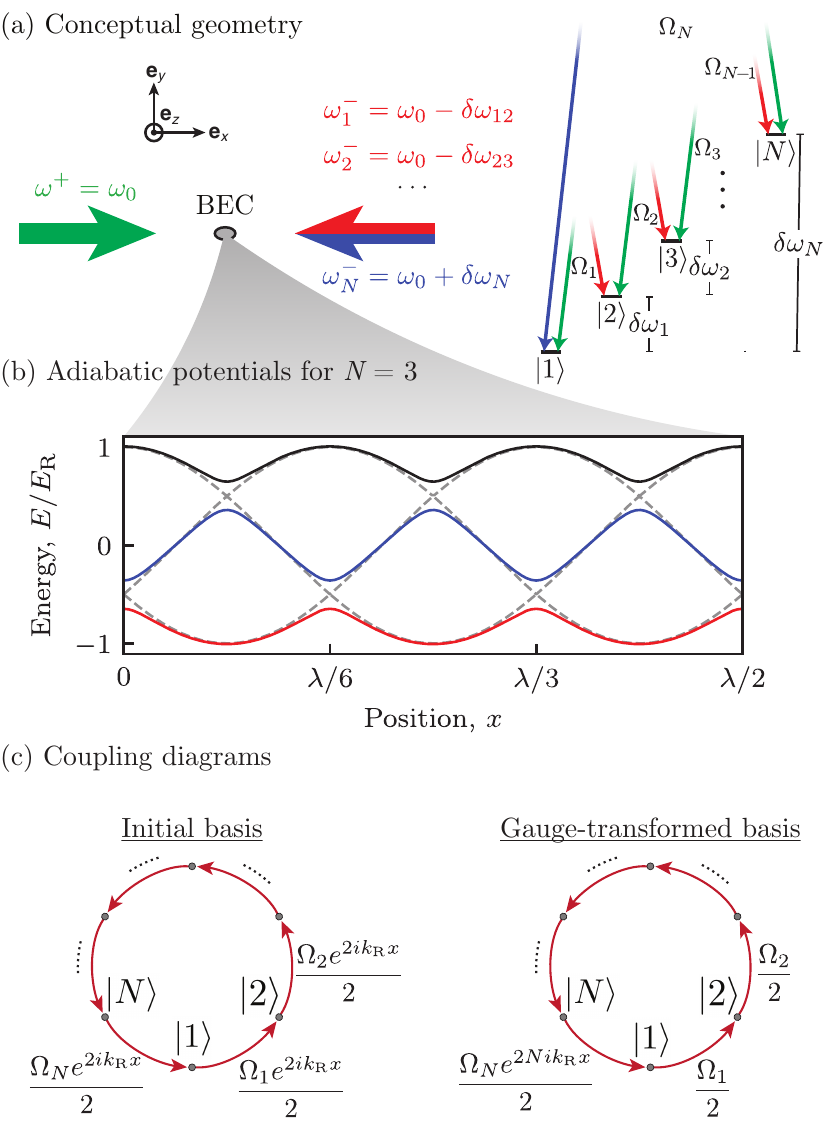}
\caption{Experimental and conceptional schematic.
\textbf{a}. Experimental geometry and level diagram.  A BEC is illuminated by a pair of laser beams that complete two photon Raman transitions between $N$ internal states. 
\textbf{b}. Adiabatic potentials illustrating the spatial subdivision of this lattice computed for $N=3$.  The gray dashed curves were computed for $\hbar {\boldsymbol \Omega} = (1.0, 1.0, 1.0)\shorttimes \Er$ and ${\boldsymbol \delta} = \mathbf{0}$; the colored curves were computed for $\hbar {\boldsymbol \Omega} = (1.25, 0.75, 1.0)\shorttimes \Er$ and  ${\boldsymbol \delta} = \mathbf{0}$.
\textbf{c}. Spin-state coupling diagrams showing the laser induced phase factors on each coupled link (left) transferred to a single link (right) after the spin-dependent gauge transformation.
}
\label{fig:setup}
\end{center}
\end{figure}

{\it Proposal} Our lattice derives from $N$ cyclically coupled internal atomic states~\cite{Campbell2011} labeled by $\ket{n}$, shown in Fig.~\ref{fig:setup}a.
Two-photon Raman transitions from lasers counter-propagating along $\ex$ link consecutive states via $[\hbar\Omega_{n}\exp(2 i \kr \hat x)/2]\ket{n+1}\bra{n}$, where $\Er = \hbar^2 \kr^2/2m$ is the single photon recoil energy derived from the ``Raman lasers'' with wavelength $\lambda$ illuminating atoms with mass $m$.
The resultant light-matter interaction term 
\begin{align}
\frac{\hat V(\hat x)}{\hbar} &=\sum_{n=1}^N \frac{\Omega_{n}}{2}e ^{2 i \kr \hat x} \ket{n\!+\!1}\bra{n} + {\rm H.c} \label{eq:HamiltonianLM}
\end{align}
and state-dependent energy shifts $\hat \Delta/\hbar = \sum_n \delta_n \ket{n}\bra{n}$ are manifestly invariant under discrete spatial translations that give an apparent $\lambda/2$ unit cell.
This lattice's true nature is first evidenced by the adiabatic potentials resulting from diagonalizing $\hat V(\hat x) + \hat \Delta$, as plotted in Fig.~\ref{fig:setup}b for $N=3$.
These potentials repeat three times within the purported $\lambda/2$ unit cell, suggesting an $N$-fold reduced unit cell.
This reduction is made explicit by a spin-dependent gauge transformation $\hat \Phi(\hat x) = \exp(i \sum_n 2 n \kr \hat x \ket{n}\bra{n})$ that leaves $\hat \Delta$ unchanged, but takes $\hat V(\hat x)$ into
\begin{align*}
\frac{\hat V^\prime(\hat x)}{\hbar} &= \left(\frac{\Omega_{N}}{2} e ^{2 N i \kr \hat x} \ket{1}\bra{N} + \sum_{n=1}^{N-1} \frac{\Omega_{n}}{2} \ket{n\!+\!1}\bra{n} \right) + {\rm H.c}.,
\end{align*}
revealing a unit cell~\cite{Cooper2011} with size $d=\lambda/2N$ and a $k_0 = 2 \kr N$ reciprocal lattice vector.
[Had Eq.~\eqref{eq:HamiltonianLM} been constructed with $M$ phase factors with reversed sign, the reciprocal lattice vector becomes  $k_0 = 2 \kr |N-2M|$ instead.]
The relation between $\hat V(\hat x)$ and $\hat V^\prime(\hat x)$ is graphically shown in Fig.~\ref{fig:setup}c.
Similar to 1D spin-orbit coupling experiments~\cite{Lin2011}, this gauge transformation also introduces a spatially uniform vector-gauge potential $\hat A\!=\!- \sum_n 2 \kr n \ket{n}\bra{n}$ taking $\hat k \rightarrow \hat k - \hat A$.  During the preparation of this manuscript, we learned of a proposal~\cite{Yan2018} that focuses on a morally similar lattice for ring-shaped traps using Laguerre-Gauss Raman laser modes, but notes the connection to linear geometries.

%
%

{\it Implementation} Our experiments began with nearly pure $\Rb87$ BECs in a crossed optical dipole trap~\cite{Lin2009}, with frequencies $(f_x,f_y,f_z) \approx \unit[(45, 55, 160)]{Hz}$.  
A magnetic field $B_0 = \unit[3.4031(1)]{mT}$ along $\ez$ Zeeman-split the $\ket{m_F=-1,0,+1}$ states comprising the ground state $F=1$ manifold by $\omega_Z/2\pi \approx \unit[23.9]{MHz}$.
We first coupled these states with strength $\Omega_{\rm rf}/2\pi = \unit[134.5(1)]{kHz}$, using a strong radio frequency (RF) magnetic field oscillating along $\ex$ with frequency $\omega_Z$.
This made a trio of synthetic clock states~\cite{Campbell2016a, Trypogeorgos2018, Anderson2018} denoted by $\ket{x}$, $\ket{y}$ and $\ket{z}$ with energies $\hbar\omega_{x,y,z}$.
Following the notation of Eq.~\eqref{eq:HamiltonianLM}, $\ket{x, y, {\rm or}\ z}$ could be equivalently labeled by $\ket{n=0, +1,{\rm or}\ -1}$ (modulo 3), and in what follows we will always take $n \in \{x, y, z\}$.
The resulting three energy splittings  $(\delta\omega_{x}, \delta\omega_{y}, \delta\omega_{z})/2\pi = \unit[(99.2(1), 281.6(1), 182.4(1))]{kHz}$ with $\delta \omega_n = |\omega_n - \omega_{n+1}|$,  were at least first-order insensitive to magnetic field fluctuations, rendering our experiment practically immune to the $\unit[0.1]{\upmu T}$ scale magnetic field noise in our laboratory.
These states are further immune to the 2-body spin relaxation collisions that plague hyperfine mixtures, and can be Raman-coupled using lasers detuned far from resonance as compared to the excited state hyperfine splitting~\cite{Campbell2016a}.

We coupled the synthetic clock states using a pair of cross-polarized Raman laser beams counter-propagating along $\ex$ tuned to the `magic-zero' wavelength $\lambda = \unit[790]{nm}$ where the scalar light shift vanishes~\cite{lamporesi_scattering_2010,*leonard_high-precision_2015}.  
The Raman beam traveling along $+\ex$ had frequency~\footnote{The added $\omega_Z$ in this expression as compared to Fig.~\ref{fig:setup}a is removed by a rotating frame transformation present in our RF dressed synthetic clock states.} $\omega^+ = \omega_0 + \omega_Z$, while the beam traveling along $-\ex$ carried frequencies $\omega^-_n = \omega_0 \pm (\delta\omega_n + \delta_n)$.
In what follows we maintain the detunings $\delta_j \approx 0$.   
The $\pm$ is selected as indicated in Fig.~\ref{fig:setup}a, such that the final `down-going' transition has the opposite frequency shift to impart the same phase factor as the `up-going' transitions.
Changing the sign on any $\delta\omega_n$ retains the Raman resonance condition, but inverts the sign of the associated phase factor in Eq.~\eqref{eq:HamiltonianLM}, generally increasesing the unit cell size.
The coupling strength of each Raman transition $\hbar\Omega_n\lesssim \Er$ from state $\ket{n}$ to $\ket{n+1}$
was far smaller than the spacing between the $\ket{x,y,z}$ states.  
This simultaneously ensured the validity of the rotating-wave approximation and rendered negligible second-order energy shifts due to off-resonant coupling to other transitions.

In the following experiments, we prepared the BEC in any of the three $\ket{x,y,z}$ states~\cite{Trypogeorgos2018} before applying any additional coupling fields.
We measured the final state by first abruptly turning off the Raman lasers and the dipole trap, thereby projecting onto the $\ket{x,y,z}$ states and free momentum states.
We then adiabatically transformed the $\ket{x,y,z}$ states back to the standard $\ket{m_F}$ states.
During the following $\unit[21]{ms}$ time-of-flight (TOF) we Stern-Gerlach separated these states and absorption imaged the resulting spin-resolved momentum distribution.

%
%

\begin{figure}[t]
\begin{center}
\includegraphics[width=3.3in]{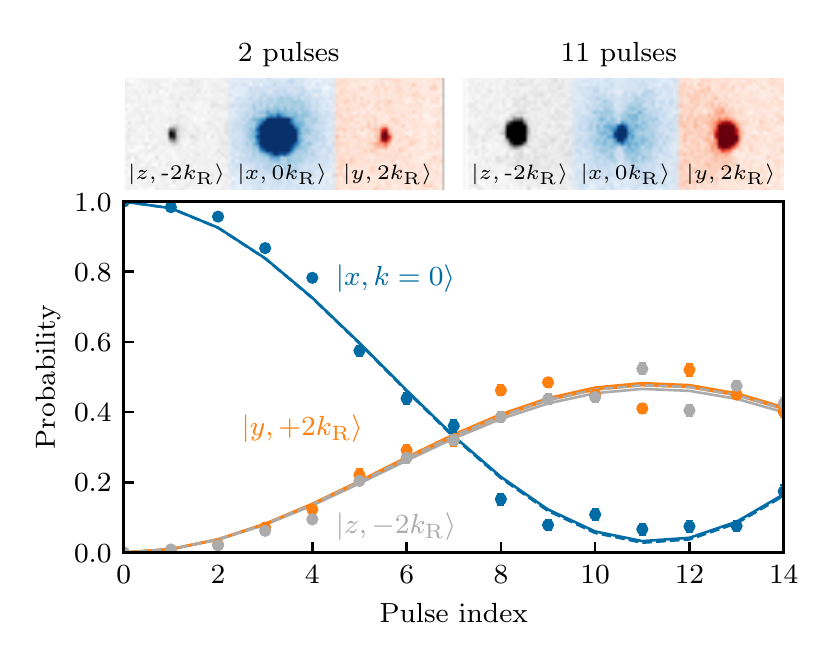}
\caption{Stroboscopic Talbot evolution starting in $\ket{x}$.
During each pulse the Raman lasers were applied for $\unit[50]{\upmu s}$ and then removed for $\unit[16.8]{\upmu s}$, giving a total pulse duration close to $h/(4\Er)$.
Top: Absorption images showing the initial $k=0$ state and diffracted orders for 2 and 11 pulses, respectively.
The symbols plot the population $\ket{x}$, and $\ket{y}$, and $\ket{z}$, colored blue, orange and gray respectively.
The solid curves depict our full lattice model using calibrated couplings $\hbar{\boldsymbol \Omega} = (0.57(1), 0.58(1), 0.58(1))\shorttimes \Er$, and detunings $\hbar{\boldsymbol \delta} = (-0.03(1), 0, 0)\shorttimes \Er$.
The detuning of the initial state $\delta_x$ was the only fit parameter.
The dashed curves plots the prediction of the simple lattice model with a depth $\hbar(\Omega_{x}+\Omega_{z}) = 1.15(2)\Er$, and the same $-0.03 \Er$ energy shift of the initial state.
}
\label{fig:Talbot}
\end{center}
\end{figure}

{\it Unitary evolution} The three gray dashed adiabatic potentials shown in Fig.~\ref{fig:setup}b rightly suggest that in the simple case of uniform coupling $\Omega_n = \bar\Omega$ and zero detuning ${\boldsymbol \delta} = {\bf 0}$, the lattice decomposes into three independent sinusoidal lattices each with depth $2\hbar\bar\Omega$ obtained by diagonalizing $\hat V^\prime(\hat x)$.
We confirmed this picture by following the unitary evolution of a BEC suddenly exposed to all three Raman fields simultaneously and observed diffraction into discrete momentum orders spaced by $6 \hbar \kr$ within each final spin state, and, as shown in the top panel of Fig.~\ref{fig:Talbot}, offset by  $\pm 2 \hbar \kr$ in the $\ket{y}$ and $\ket{z}$ states respectively.
For any initial spin state, the dynamics of these orders individually were indistinguishable from the $2 \hbar \kr n$ orders diffracting off a conventional 1D optical lattice.
We enhanced the diffraction from our comparatively shallow lattice by pulsing it repeatedly~\cite{Herold2012}: alternating between periods of evolution with and without the lattice present allows a state initially in $\ket{k=0}$ to acquire far more population in $\ket{k=\pm 2 \kr}$ than from a single uninterrupted pulse of any duration.
Figure~\ref{fig:Talbot} shows the resulting evolution for a system initially in $\ket{x}$; the solid curves depict the prediction of our full lattice model using independently calibrated couplings (see caption).
For weak Raman coupling such as ours, the matrix elements directly coupling the initial state dominate the dynamics.
The dashed curves depict the prediction of a simple 1D lattice with depth $\hbar (\Omega_{x} + \Omega_{z}) = 1.15(2)\Er$.
This overall agreement validates our underlying model, but is insufficient to demonstrate the expanded Brillouin zone (BZ).

%
%

\begin{figure}[t]
\begin{center}
\includegraphics{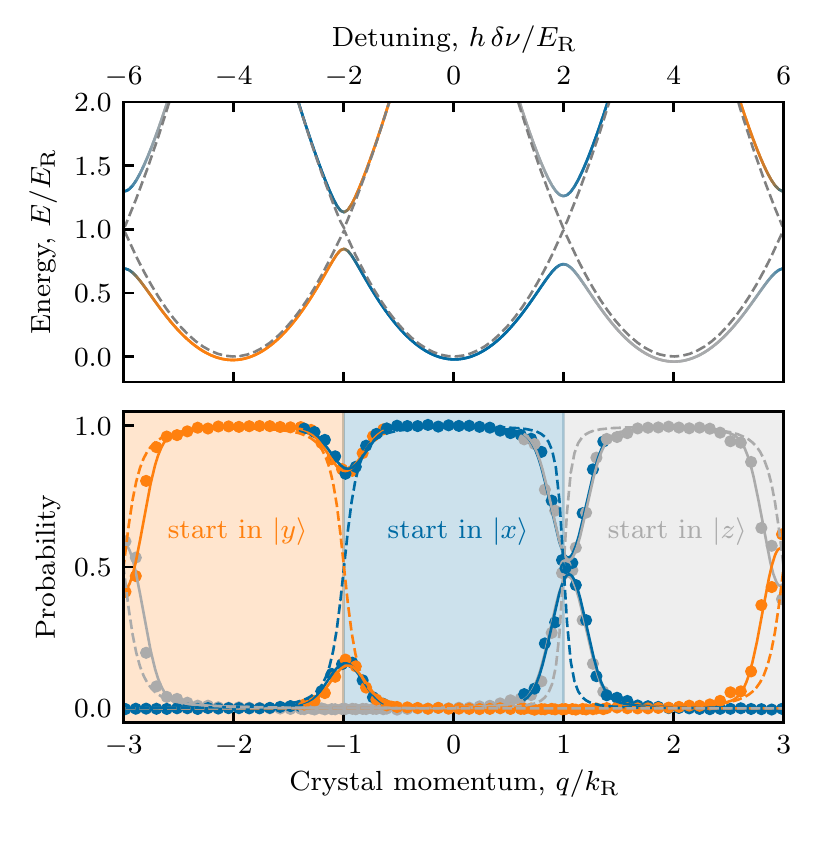}
\caption{Expanded BZ.  
The top axis shows the frequency shift $\delta\nu$ used to affect a moving lattice that populated the desired $\hbar q$ state.
Top: Computed band structure.  Each solid curve is shaded in accordance to the population in the $\ket{x}$ (blue), $\ket{y}$ (orange), $\ket{z}$ (gray) states.  
The dashed curves depict the bare free particle dispersion absent Raman coupling.
Bottom: Spin composition of the lowest band.
Different regions of the BZ were explored by starting in $\ket{x}$, $\ket{y}$, and $\ket{z}$.
The dashed curves denote the spin composition of the lowest band, while the solid curves plot the outcome of a full simulation of our experimental protocol.  We determine the Raman coupling strengths to be $\hbar {\boldsymbol \Omega} = (0.29(1), 0.61(1), 0.54(1)) \, \Er$ and detunings $\hbar \vert \delta_{x,y,z} \vert \leq 0.07(1) \Er$.
}
\label{fig:EnlargedBZ}
\end{center}
\end{figure}

{\it Enlarged BZ}  We experimentally resolved the enlarged BZ by measuring the spin-state composition in the lowest band as a function of the crystal momentum $\hbar q$ and show that it repeats every $6\hbar\kr$ rather than every $2\hbar\kr$ as would be expected for a $\lambda/2$ lattice period.
We used the narrow momentum distribution of a BEC to probe individual crystal momentum states, and rather than accelerating the BEC, we ramped on the lattice in $\unit[200]{\upmu s}$, adiabatically loading BECs at rest in the lab frame into non-zero crystal momentum states of a moving lattice.
We brought the lattice into motion~\cite{browaeys_transport_2005,peik_bloch_1997} by detuning one of the two Raman lasers by $\delta \nu$, giving a crystal momentum of  $q / \kr = h \, \delta \nu / (4 \Er)$ in the lattice's rest frame.
After a brief $\unit[50]{\upmu s}$ hold in the moving lattice, we measured the spin-resolved momentum distribution.
As shown in Fig.~\ref{fig:EnlargedBZ}(top), the lowest band contains three local minima near $q = -2 \kr$, $0$, or $2 \kr$, predominantly derived from the $\ket{y}$, $\ket{x}$, or $\ket{z}$ state respectively.
The first excited band approaches the ground band at avoided crossings between these minima, rendering our lattice turn-on non-adiabatic in their vicinity.
Accordingly, we accessed the entirety of the expanded BZ in a piecewise manner: for each of the three initial spin states, we applied the above method to focus on a single $2\hbar\kr$ interval.

We operated in a regime of imbalanced coupling (see caption), where the adiabatic potential cannot be decomposed into independent sinusoids.
Figure~\ref{fig:EnlargedBZ}(bottom) shows the measured occupation probability in each of the $\ket{x,y,z}$ states, immediately exposing the enlarged BZ.
The dashed curves depict the occupation probabilities in the ground band for our parameters, which agree with our data away from regimes of non-adiabatic loading.
The solid curves are the result of a numerical simulation of our loading procedure including all non-adiabatic effects which are in near-perfect agreement with our measurements.
The differing population ratios in the vicinity of the avoided crossings result from the asymmetric Raman coupling.

%
%

\begin{figure}[t]
\begin{center}
\includegraphics{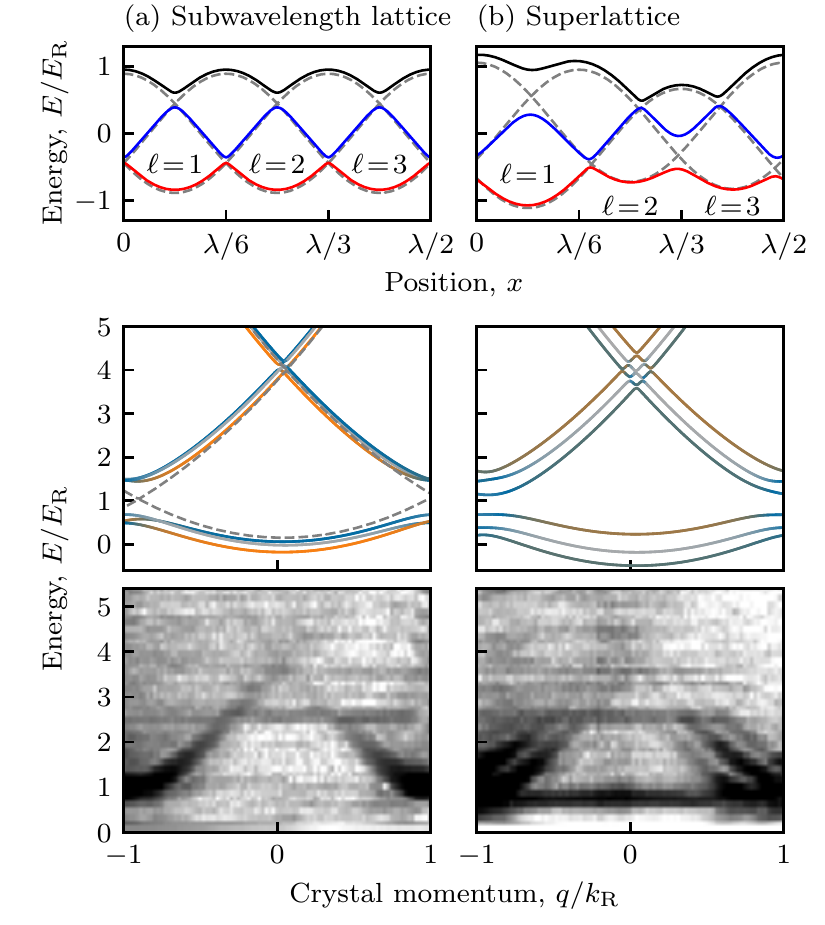}
\caption{Superlattice.
Panels (a) and (b) depict $\Omega^{\rm rf}_n = 0$ (Raman-only) and with $\hbar\Omega^{\rm rf}_x =  0.7 \Er$ (Raman and RF), respectively.
In both cases the Raman coupling strengths were $\hbar {\boldsymbol \Omega} = (0.98, 0.82, 0.87)\shorttimes \Er$ and the detunings were  $\hbar {\boldsymbol \delta} = (0.08, 0.15, 0.08)\shorttimes \Er$.
Top: computed adiabatic potentials with and without RF, showing the formation of a superlattice with RF.  The RF phase determines the energy of the sublattice sites, and was selected to be $\phi_x = \pi/4$ for this simulation.
Middle: computed band structure for the experimental parameters (shaded per Fig.~\ref{fig:EnlargedBZ}).
Bottom: measured PSDs, obtained averaging five time series. 
In each map, the intensity reflects the combined PSD associated with each initial crystal momentum state.
For the Raman-only data in (a) we started in the $\ket{z}$ state only, while for the Raman and RF data in (b) we increased the signal to noise ratio by combining data starting in $\ket{x}$, $\ket{y}$ and $\ket{z}$.  The horizontal structures present in both data sets around $2.5\Er$ result from technical noise present in the laboratory.
}
\label{fig:PSD}
\end{center}
\end{figure}

{\it Superlattice} We conclude by describing how to gain individual control over the energy-minima of the $N$ sub-lattice sites discussed above, essentially combining $N$ reduced unit cells into a superlattice with $N$ sites.
We demonstrate this principle by creating a tunable triple-well lattice.  
We introduce superlattice terms with conventional $\lambda/2$ periodicity by adding a spatially homogenous coupling with strength $\Omega^{\rm rf}_n$ and phase $\phi^{\rm rf}_n$ to the links in Eq.~\ref{eq:HamiltonianLM}, giving combined matrix elements such as $[\Omega_n\exp(2 i \kr \hat x) + \Omega^{\rm rf}_n \exp(i \phi^{\rm rf}_n)]/2$.
For weak RF, the most significant effect of even one such RF coupling is to shift the energies within the unit cell as shown in the top panels of Fig.~\ref{fig:PSD}, where column (a) denotes the $\lambda/2$ unit cell case with no RF and column (b) shows the impact of RF.
As in Fig.~\ref{fig:setup}b, the dashed curves show the three uncoupled sinusoids present for uniform detuning and Raman coupling, that we now enumerate with $\ell$, ranging from $1$ to $N$.  
At lowest order RF shifts these curves by an energy $\hbar\Omega^{\rm rf}_n\cos(2\pi\ell/N + \phi^{\rm rf}_n)/N$, and at higher order it introduces new transition matrix elements between the adiabatic potentials.
Just as the familiar bipartite (double-well) Su-Schrieffer-Heeger model~\cite{Su1979} has a pair of low-energy bands, this $N$-partite lattice has $N$ low-energy bands each occupying the initial (not enlarged) BZ.

We demonstrated this concept by adding one RF coupling field to our Raman lattice and directly verified the formation of the superlattice potential using Fourier transform spectroscopy~\cite{Valdes-Curiel2017}.
Rather than working with Bose-condensed atoms with momentum width $\Delta k \ll \kr$, we used a non-condensed cloud with temperature $T \approx \unit[200]{nK}$ and $T/T_C\approx1.1$, where $T_C$ is the BEC transition temperature.  This allowed us to simultaneously sample a range of momentum states spanning the whole BZ.
We simultaneously pulsed the cyclical Raman couplings and a single RF coupling $\Omega^{\rm rf}_x$ linking $\ket{x}$ and $\ket{y}$. We then measured the resulting time-evolving spin-resolved momentum distributions for $\unit[2]{ms}$, giving a $\approx \unit[0.5]{kHz}$ frequency resolution.

Figure~\ref{fig:PSD} shows the associated power spectral densities (PSDs) both without (left) and with (right) RF coupling, expressed in the initial (not enlarged) BZ.
The Raman-only data (left) are dominated by a single difference frequency associated with the Raman lattice's splitting; here the degenerate spectra associated with the enlarged BZ have been translated and lie atop each other.
The addition of RF coupling (right) lifted this degeneracy and produced three sub-bands, each associated with a single site of the $3$-partite lattice.

%
%

Our techniques are readily extensible to higher spatial dimensions~\cite{Cooper2012}: in 2D, Kagome lattices can be generated using the same three internal states used here; and in 3D pyrochlore-type lattices can be assembled using four internal states.
The latter is of particular interest as it is a candidate lattice for realizing non-Abelian topological spin and charge pumps~\cite{Lu2018} derived from the second Chern number~\cite{Sugawa2018}.
Closer to home, the specific three-site superlattice demonstrated here is ideally suited for assembling gauge fields without spatial gradients~\cite{Kennedy2013}.

\begin{acknowledgments}
This work was partially supported by the AFOSRs Quantum Matter MURI, NIST, and the NSF through the PFC at the JQI.
\end{acknowledgments}

\bibliography{FractionalLattice,FractionalLatticeAdditions}

\begin{thebibliography}{35}%
\makeatletter
\providecommand \@ifxundefined [1]{%
 \@ifx{#1\undefined}
}%
\providecommand \@ifnum [1]{%
 \ifnum #1\expandafter \@firstoftwo
 \else \expandafter \@secondoftwo
 \fi
}%
\providecommand \@ifx [1]{%
 \ifx #1\expandafter \@firstoftwo
 \else \expandafter \@secondoftwo
 \fi
}%
\providecommand \natexlab [1]{#1}%
\providecommand \enquote  [1]{``#1''}%
\providecommand \bibnamefont  [1]{#1}%
\providecommand \bibfnamefont [1]{#1}%
\providecommand \citenamefont [1]{#1}%
\providecommand \href@noop [0]{\@secondoftwo}%
\providecommand \href [0]{\begingroup \@sanitize@url \@href}%
\providecommand \@href[1]{\@@startlink{#1}\@@href}%
\providecommand \@@href[1]{\endgroup#1\@@endlink}%
\providecommand \@sanitize@url [0]{\catcode `\\12\catcode `\$12\catcode
  `\&12\catcode `\#12\catcode `\^12\catcode `\_12\catcode `\%12\relax}%
\providecommand \@@startlink[1]{}%
\providecommand \@@endlink[0]{}%
\providecommand \url  [0]{\begingroup\@sanitize@url \@url }%
\providecommand \@url [1]{\endgroup\@href {#1}{\urlprefix }}%
\providecommand \urlprefix  [0]{URL }%
\providecommand \Eprint [0]{\href }%
\providecommand \doibase [0]{http://dx.doi.org/}%
\providecommand \selectlanguage [0]{\@gobble}%
\providecommand \bibinfo  [0]{\@secondoftwo}%
\providecommand \bibfield  [0]{\@secondoftwo}%
\providecommand \translation [1]{[#1]}%
\providecommand \BibitemOpen [0]{}%
\providecommand \bibitemStop [0]{}%
\providecommand \bibitemNoStop [0]{.\EOS\space}%
\providecommand \EOS [0]{\spacefactor3000\relax}%
\providecommand \BibitemShut  [1]{\csname bibitem#1\endcsname}%
\let\auto@bib@innerbib\@empty
\bibitem [{\citenamefont {Greiner}\ \emph {et~al.}(2002)\citenamefont
  {Greiner}, \citenamefont {Mandel}, \citenamefont {Esslinger}, \citenamefont
  {Hansch},\ and\ \citenamefont {Bloch}}]{Greiner2002}%
  \BibitemOpen
  \bibfield  {author} {\bibinfo {author} {\bibfnamefont {M.}~\bibnamefont
  {Greiner}}, \bibinfo {author} {\bibfnamefont {O.}~\bibnamefont {Mandel}},
  \bibinfo {author} {\bibfnamefont {T.}~\bibnamefont {Esslinger}}, \bibinfo
  {author} {\bibfnamefont {T.~W.}\ \bibnamefont {Hansch}}, \ and\ \bibinfo
  {author} {\bibfnamefont {I.}~\bibnamefont {Bloch}},\ }\href@noop {}
  {\bibfield  {journal} {\bibinfo  {journal} {Nature}\ }\textbf {\bibinfo
  {volume} {415}},\ \bibinfo {pages} {39} (\bibinfo {year} {2002})}\BibitemShut
  {NoStop}%
\bibitem [{\citenamefont {Wu}\ \emph {et~al.}(2016)\citenamefont {Wu},
  \citenamefont {Zhang}, \citenamefont {Sun}, \citenamefont {Xu}, \citenamefont
  {Wang}, \citenamefont {Ji}, \citenamefont {Deng}, \citenamefont {Chen},
  \citenamefont {Liu},\ and\ \citenamefont {Pan}}]{Wu2016}%
  \BibitemOpen
  \bibfield  {author} {\bibinfo {author} {\bibfnamefont {Z.}~\bibnamefont
  {Wu}}, \bibinfo {author} {\bibfnamefont {L.}~\bibnamefont {Zhang}}, \bibinfo
  {author} {\bibfnamefont {W.}~\bibnamefont {Sun}}, \bibinfo {author}
  {\bibfnamefont {X.-T.}\ \bibnamefont {Xu}}, \bibinfo {author} {\bibfnamefont
  {B.-Z.}\ \bibnamefont {Wang}}, \bibinfo {author} {\bibfnamefont {S.-C.}\
  \bibnamefont {Ji}}, \bibinfo {author} {\bibfnamefont {Y.}~\bibnamefont
  {Deng}}, \bibinfo {author} {\bibfnamefont {S.}~\bibnamefont {Chen}}, \bibinfo
  {author} {\bibfnamefont {X.-J.}\ \bibnamefont {Liu}}, \ and\ \bibinfo
  {author} {\bibfnamefont {J.-W.}\ \bibnamefont {Pan}},\ }\href {\doibase
  10.1126/science.aaf6689} {\bibfield  {journal} {\bibinfo  {journal}
  {Science}\ }\textbf {\bibinfo {volume} {354}},\ \bibinfo {pages} {83}
  (\bibinfo {year} {2016})}\BibitemShut {NoStop}%
\bibitem [{\citenamefont {Bloom}\ \emph {et~al.}(2014)\citenamefont {Bloom},
  \citenamefont {Nicholson}, \citenamefont {Williams}, \citenamefont
  {Campbell}, \citenamefont {Bishof}, \citenamefont {Zhang}, \citenamefont
  {Zhang}, \citenamefont {Bromley},\ and\ \citenamefont {Ye}}]{Bloom2014}%
  \BibitemOpen
  \bibfield  {author} {\bibinfo {author} {\bibfnamefont {B.~J.}\ \bibnamefont
  {Bloom}}, \bibinfo {author} {\bibfnamefont {T.~L.}\ \bibnamefont
  {Nicholson}}, \bibinfo {author} {\bibfnamefont {J.~R.}\ \bibnamefont
  {Williams}}, \bibinfo {author} {\bibfnamefont {S.~L.}\ \bibnamefont
  {Campbell}}, \bibinfo {author} {\bibfnamefont {M.}~\bibnamefont {Bishof}},
  \bibinfo {author} {\bibfnamefont {X.}~\bibnamefont {Zhang}}, \bibinfo
  {author} {\bibfnamefont {W.}~\bibnamefont {Zhang}}, \bibinfo {author}
  {\bibfnamefont {S.~L.}\ \bibnamefont {Bromley}}, \ and\ \bibinfo {author}
  {\bibfnamefont {J.}~\bibnamefont {Ye}},\ }\href@noop {} {\bibfield  {journal}
  {\bibinfo  {journal} {Nature}\ }\textbf {\bibinfo {volume} {506}},\ \bibinfo
  {pages} {71 EP } (\bibinfo {year} {2014})}\BibitemShut {NoStop}%
\bibitem [{\citenamefont {Gupta}\ \emph {et~al.}(1996)\citenamefont {Gupta},
  \citenamefont {McClelland}, \citenamefont {Marte},\ and\ \citenamefont
  {Celotta}}]{gupta_raman-induced_1996}%
  \BibitemOpen
  \bibfield  {author} {\bibinfo {author} {\bibfnamefont {R.}~\bibnamefont
  {Gupta}}, \bibinfo {author} {\bibfnamefont {J.~J.}\ \bibnamefont
  {McClelland}}, \bibinfo {author} {\bibfnamefont {P.}~\bibnamefont {Marte}}, \
  and\ \bibinfo {author} {\bibfnamefont {R.~J.}\ \bibnamefont {Celotta}},\
  }\href {\doibase 10.1103/PhysRevLett.76.4689} {\bibfield  {journal} {\bibinfo
   {journal} {Physical Review Letters}\ }\textbf {\bibinfo {volume} {76}},\
  \bibinfo {pages} {4689} (\bibinfo {year} {1996})}\BibitemShut {NoStop}%
\bibitem [{\citenamefont {Lundblad}\ \emph {et~al.}(2008)\citenamefont
  {Lundblad}, \citenamefont {Lee}, \citenamefont {Spielman}, \citenamefont
  {Brown}, \citenamefont {Phillips},\ and\ \citenamefont
  {Porto}}]{Lundblad2008}%
  \BibitemOpen
  \bibfield  {author} {\bibinfo {author} {\bibfnamefont {N.}~\bibnamefont
  {Lundblad}}, \bibinfo {author} {\bibfnamefont {P.~J.}\ \bibnamefont {Lee}},
  \bibinfo {author} {\bibfnamefont {I.~B.}\ \bibnamefont {Spielman}}, \bibinfo
  {author} {\bibfnamefont {B.~L.}\ \bibnamefont {Brown}}, \bibinfo {author}
  {\bibfnamefont {W.~D.}\ \bibnamefont {Phillips}}, \ and\ \bibinfo {author}
  {\bibfnamefont {J.~V.}\ \bibnamefont {Porto}},\ }\href {\doibase
  10.1103/PhysRevLett.100.150401} {\bibfield  {journal} {\bibinfo  {journal}
  {Physical Review Letters}\ }\textbf {\bibinfo {volume} {100}},\ \bibinfo
  {pages} {150401} (\bibinfo {year} {2008})}\BibitemShut {NoStop}%
\bibitem [{\citenamefont {Yi}\ \emph {et~al.}(2008)\citenamefont {Yi},
  \citenamefont {Daley}, \citenamefont {Pupillo},\ and\ \citenamefont
  {Zoller}}]{Yi2008}%
  \BibitemOpen
  \bibfield  {author} {\bibinfo {author} {\bibfnamefont {W.}~\bibnamefont
  {Yi}}, \bibinfo {author} {\bibfnamefont {A.~J.}\ \bibnamefont {Daley}},
  \bibinfo {author} {\bibfnamefont {G.}~\bibnamefont {Pupillo}}, \ and\
  \bibinfo {author} {\bibfnamefont {P.}~\bibnamefont {Zoller}},\ }\href
  {http://stacks.iop.org/1367-2630/10/073015} {\bibfield  {journal} {\bibinfo
  {journal} {New Journal of Physics}\ }\textbf {\bibinfo {volume} {10}},\
  \bibinfo {pages} {073015 (22pp)} (\bibinfo {year} {2008})}\BibitemShut
  {NoStop}%
\bibitem [{\citenamefont {Nascimbene}\ \emph {et~al.}(2015)\citenamefont
  {Nascimbene}, \citenamefont {Goldman}, \citenamefont {Cooper},\ and\
  \citenamefont {Dalibard}}]{Nascimbene2015}%
  \BibitemOpen
  \bibfield  {author} {\bibinfo {author} {\bibfnamefont {S.}~\bibnamefont
  {Nascimbene}}, \bibinfo {author} {\bibfnamefont {N.}~\bibnamefont {Goldman}},
  \bibinfo {author} {\bibfnamefont {N.~R.}\ \bibnamefont {Cooper}}, \ and\
  \bibinfo {author} {\bibfnamefont {J.}~\bibnamefont {Dalibard}},\ }\href
  {\doibase 10.1103/PhysRevLett.115.140401} {\bibfield  {journal} {\bibinfo
  {journal} {Phys. Rev. Lett.}\ }\textbf {\bibinfo {volume} {115}},\ \bibinfo
  {pages} {140401} (\bibinfo {year} {2015})}\BibitemShut {NoStop}%
\bibitem [{\citenamefont {{\L}{\k a}cki}\ \emph {et~al.}(2016)\citenamefont
  {{\L}{\k a}cki}, \citenamefont {Baranov}, \citenamefont {Pichler},\ and\
  \citenamefont {Zoller}}]{Lacki2016}%
  \BibitemOpen
  \bibfield  {author} {\bibinfo {author} {\bibfnamefont {M.}~\bibnamefont
  {{\L}{\k a}cki}}, \bibinfo {author} {\bibfnamefont {M.~A.}\ \bibnamefont
  {Baranov}}, \bibinfo {author} {\bibfnamefont {H.}~\bibnamefont {Pichler}}, \
  and\ \bibinfo {author} {\bibfnamefont {P.}~\bibnamefont {Zoller}},\ }\href
  {\doibase 10.1103/PhysRevLett.117.233001} {\bibfield  {journal} {\bibinfo
  {journal} {Phys. Rev. Lett.}\ }\textbf {\bibinfo {volume} {117}},\ \bibinfo
  {pages} {233001} (\bibinfo {year} {2016})}\BibitemShut {NoStop}%
\bibitem [{\citenamefont {Jendrzejewski}\ \emph {et~al.}(2016)\citenamefont
  {Jendrzejewski}, \citenamefont {Eckel}, \citenamefont {Tiecke}, \citenamefont
  {Juzeli\ifmmode~\bar{u}\else \={u}\fi{}nas}, \citenamefont {Campbell},
  \citenamefont {Jiang},\ and\ \citenamefont {Gorshkov}}]{Jendrzejewski2016}%
  \BibitemOpen
  \bibfield  {author} {\bibinfo {author} {\bibfnamefont {F.}~\bibnamefont
  {Jendrzejewski}}, \bibinfo {author} {\bibfnamefont {S.}~\bibnamefont
  {Eckel}}, \bibinfo {author} {\bibfnamefont {T.~G.}\ \bibnamefont {Tiecke}},
  \bibinfo {author} {\bibfnamefont {G.}~\bibnamefont
  {Juzeli\ifmmode~\bar{u}\else \={u}\fi{}nas}}, \bibinfo {author}
  {\bibfnamefont {G.~K.}\ \bibnamefont {Campbell}}, \bibinfo {author}
  {\bibfnamefont {L.}~\bibnamefont {Jiang}}, \ and\ \bibinfo {author}
  {\bibfnamefont {A.~V.}\ \bibnamefont {Gorshkov}},\ }\href {\doibase
  10.1103/PhysRevA.94.063422} {\bibfield  {journal} {\bibinfo  {journal} {Phys.
  Rev. A}\ }\textbf {\bibinfo {volume} {94}},\ \bibinfo {pages} {063422}
  (\bibinfo {year} {2016})}\BibitemShut {NoStop}%
\bibitem [{\citenamefont {Wang}\ \emph {et~al.}(2018)\citenamefont {Wang},
  \citenamefont {Subhankar}, \citenamefont {Bienias}, \citenamefont {Lacki},
  \citenamefont {Tsui}, \citenamefont {Baranov}, \citenamefont {Gorshkov},
  \citenamefont {Zoller}, \citenamefont {Porto},\ and\ \citenamefont
  {Rolston}}]{wang_dark_2018}%
  \BibitemOpen
  \bibfield  {author} {\bibinfo {author} {\bibfnamefont {Y.}~\bibnamefont
  {Wang}}, \bibinfo {author} {\bibfnamefont {S.}~\bibnamefont {Subhankar}},
  \bibinfo {author} {\bibfnamefont {P.}~\bibnamefont {Bienias}}, \bibinfo
  {author} {\bibfnamefont {M.}~\bibnamefont {Lacki}}, \bibinfo {author}
  {\bibfnamefont {T.-C.}\ \bibnamefont {Tsui}}, \bibinfo {author}
  {\bibfnamefont {M.}~\bibnamefont {Baranov}}, \bibinfo {author} {\bibfnamefont
  {A.}~\bibnamefont {Gorshkov}}, \bibinfo {author} {\bibfnamefont
  {P.}~\bibnamefont {Zoller}}, \bibinfo {author} {\bibfnamefont
  {J.}~\bibnamefont {Porto}}, \ and\ \bibinfo {author} {\bibfnamefont
  {S.}~\bibnamefont {Rolston}},\ }\href@noop {} {\bibfield  {journal} {\bibinfo
   {journal} {Physical Review Letters}\ }\textbf {\bibinfo {volume} {120}},\
  \bibinfo {pages} {083601} (\bibinfo {year} {2018})}\BibitemShut {NoStop}%
\bibitem [{\citenamefont {Weitz}\ \emph {et~al.}(2004)\citenamefont {Weitz},
  \citenamefont {Cennini}, \citenamefont {Ritt},\ and\ \citenamefont
  {Geckeler}}]{weitz_optical_2004}%
  \BibitemOpen
  \bibfield  {author} {\bibinfo {author} {\bibfnamefont {M.}~\bibnamefont
  {Weitz}}, \bibinfo {author} {\bibfnamefont {G.}~\bibnamefont {Cennini}},
  \bibinfo {author} {\bibfnamefont {G.}~\bibnamefont {Ritt}}, \ and\ \bibinfo
  {author} {\bibfnamefont {C.}~\bibnamefont {Geckeler}},\ }\href {\doibase
  10.1103/PhysRevA.70.043414} {\bibfield  {journal} {\bibinfo  {journal}
  {Physical Review A}\ }\textbf {\bibinfo {volume} {70}},\ \bibinfo {pages}
  {043414} (\bibinfo {year} {2004})}\BibitemShut {NoStop}%
\bibitem [{\citenamefont {Zhang}\ \emph {et~al.}(2005)\citenamefont {Zhang},
  \citenamefont {Morrow}, \citenamefont {Berman},\ and\ \citenamefont
  {Raithel}}]{Zhang2005}%
  \BibitemOpen
  \bibfield  {author} {\bibinfo {author} {\bibfnamefont {R.}~\bibnamefont
  {Zhang}}, \bibinfo {author} {\bibfnamefont {N.~V.}\ \bibnamefont {Morrow}},
  \bibinfo {author} {\bibfnamefont {P.~R.}\ \bibnamefont {Berman}}, \ and\
  \bibinfo {author} {\bibfnamefont {G.}~\bibnamefont {Raithel}},\ }\href@noop
  {} {\bibfield  {journal} {\bibinfo  {journal} {Physical Review A}\ }\textbf
  {\bibinfo {volume} {72}},\ \bibinfo {pages} {043409} (\bibinfo {year}
  {2005})}\BibitemShut {NoStop}%
\bibitem [{\citenamefont {Ritt}\ \emph {et~al.}(2006)\citenamefont {Ritt},
  \citenamefont {Geckeler}, \citenamefont {Salger}, \citenamefont {Cennini},\
  and\ \citenamefont {Weitz}}]{Ritt2006}%
  \BibitemOpen
  \bibfield  {author} {\bibinfo {author} {\bibfnamefont {G.}~\bibnamefont
  {Ritt}}, \bibinfo {author} {\bibfnamefont {C.}~\bibnamefont {Geckeler}},
  \bibinfo {author} {\bibfnamefont {T.}~\bibnamefont {Salger}}, \bibinfo
  {author} {\bibfnamefont {G.}~\bibnamefont {Cennini}}, \ and\ \bibinfo
  {author} {\bibfnamefont {M.}~\bibnamefont {Weitz}},\ }\href@noop {}
  {\bibfield  {journal} {\bibinfo  {journal} {Physical Review A}\ }\textbf
  {\bibinfo {volume} {74}},\ \bibinfo {pages} {063622} (\bibinfo {year}
  {2006})}\BibitemShut {NoStop}%
\bibitem [{\citenamefont {Haske}\ \emph {et~al.}(2007)\citenamefont {Haske},
  \citenamefont {Chen}, \citenamefont {Hales}, \citenamefont {Dong},
  \citenamefont {Barlow}, \citenamefont {Marder},\ and\ \citenamefont
  {Perry}}]{Haske2007}%
  \BibitemOpen
  \bibfield  {author} {\bibinfo {author} {\bibfnamefont {W.}~\bibnamefont
  {Haske}}, \bibinfo {author} {\bibfnamefont {V.~W.}\ \bibnamefont {Chen}},
  \bibinfo {author} {\bibfnamefont {J.~M.}\ \bibnamefont {Hales}}, \bibinfo
  {author} {\bibfnamefont {W.}~\bibnamefont {Dong}}, \bibinfo {author}
  {\bibfnamefont {S.}~\bibnamefont {Barlow}}, \bibinfo {author} {\bibfnamefont
  {S.~R.}\ \bibnamefont {Marder}}, \ and\ \bibinfo {author} {\bibfnamefont
  {J.~W.}\ \bibnamefont {Perry}},\ }\href {\doibase 10.1364/OE.15.003426}
  {\bibfield  {journal} {\bibinfo  {journal} {Opt. Express}\ }\textbf {\bibinfo
  {volume} {15}},\ \bibinfo {pages} {3426} (\bibinfo {year}
  {2007})}\BibitemShut {NoStop}%
\bibitem [{\citenamefont {You}\ \emph {et~al.}(2018)\citenamefont {You},
  \citenamefont {Liao}, \citenamefont {Hemmer},\ and\ \citenamefont
  {Zubairy}}]{You2018}%
  \BibitemOpen
  \bibfield  {author} {\bibinfo {author} {\bibfnamefont {J.}~\bibnamefont
  {You}}, \bibinfo {author} {\bibfnamefont {Z.}~\bibnamefont {Liao}}, \bibinfo
  {author} {\bibfnamefont {P.~R.}\ \bibnamefont {Hemmer}}, \ and\ \bibinfo
  {author} {\bibfnamefont {M.~S.}\ \bibnamefont {Zubairy}},\ }\href {\doibase
  10.1103/PhysRevA.97.043807} {\bibfield  {journal} {\bibinfo  {journal} {Phys.
  Rev. A}\ }\textbf {\bibinfo {volume} {97}},\ \bibinfo {pages} {043807}
  (\bibinfo {year} {2018})}\BibitemShut {NoStop}%
\bibitem [{\citenamefont {Campbell}\ \emph {et~al.}(2011)\citenamefont
  {Campbell}, \citenamefont {Juzeli{\=u}nas},\ and\ \citenamefont
  {Spielman}}]{Campbell2011}%
  \BibitemOpen
  \bibfield  {author} {\bibinfo {author} {\bibfnamefont {D.~L.}\ \bibnamefont
  {Campbell}}, \bibinfo {author} {\bibfnamefont {G.}~\bibnamefont
  {Juzeli{\=u}nas}}, \ and\ \bibinfo {author} {\bibfnamefont {I.~B.}\
  \bibnamefont {Spielman}},\ }\href {\doibase 10.1103/PhysRevA.84.025602}
  {\bibfield  {journal} {\bibinfo  {journal} {Phys. Rev. A}\ }\textbf {\bibinfo
  {volume} {84}} (\bibinfo {year} {2011}),\
  10.1103/PhysRevA.84.025602}\BibitemShut {NoStop}%
\bibitem [{\citenamefont {Cooper}\ and\ \citenamefont
  {Dalibard}(2011)}]{Cooper2011}%
  \BibitemOpen
  \bibfield  {author} {\bibinfo {author} {\bibfnamefont {N.~R.}\ \bibnamefont
  {Cooper}}\ and\ \bibinfo {author} {\bibfnamefont {J.}~\bibnamefont
  {Dalibard}},\ }\href {\doibase 10.1209/0295-5075/95/66004} {\bibfield
  {journal} {\bibinfo  {journal} {EPL (Europhysics Letters)}\ }\textbf
  {\bibinfo {volume} {95}},\ \bibinfo {pages} {66004} (\bibinfo {year}
  {2011})}\BibitemShut {NoStop}%
\bibitem [{\citenamefont {Lin}\ \emph {et~al.}(2011)\citenamefont {Lin},
  \citenamefont {Jim{\'e}nez-Garc{\'\i}a},\ and\ \citenamefont
  {Spielman}}]{Lin2011}%
  \BibitemOpen
  \bibfield  {author} {\bibinfo {author} {\bibfnamefont {Y.-J.}\ \bibnamefont
  {Lin}}, \bibinfo {author} {\bibfnamefont {K.}~\bibnamefont
  {Jim{\'e}nez-Garc{\'\i}a}}, \ and\ \bibinfo {author} {\bibfnamefont {I.~B.}\
  \bibnamefont {Spielman}},\ }\href {\doibase 10.1038/nature09887} {\bibfield
  {journal} {\bibinfo  {journal} {Nature}\ }\textbf {\bibinfo {volume} {471}},\
  \bibinfo {pages} {83} (\bibinfo {year} {2011})}\BibitemShut {NoStop}%
\bibitem [{\citenamefont {Yan}\ \emph {et~al.}(2018)\citenamefont {Yan},
  \citenamefont {Zhang}, \citenamefont {Choudhury},\ and\ \citenamefont
  {Zhou}}]{Yan2018}%
  \BibitemOpen
  \bibfield  {author} {\bibinfo {author} {\bibfnamefont {Y.}~\bibnamefont
  {Yan}}, \bibinfo {author} {\bibfnamefont {S.-L.}\ \bibnamefont {Zhang}},
  \bibinfo {author} {\bibfnamefont {S.}~\bibnamefont {Choudhury}}, \ and\
  \bibinfo {author} {\bibfnamefont {Q.}~\bibnamefont {Zhou}},\ }\href@noop {}
  {\enquote {\bibinfo {title} {Emergent periodic and quasiperiodic lattices on
  surfaces of synthetic hall tori and synthetic hall cylinders},}\ } (\bibinfo
  {year} {2018}),\ \Eprint {http://arxiv.org/abs/arXiv:1810.12331}
  {arXiv:1810.12331} \BibitemShut {NoStop}%
\bibitem [{\citenamefont {Lin}\ \emph {et~al.}(2009)\citenamefont {Lin},
  \citenamefont {Perry}, \citenamefont {Compton}, \citenamefont {Spielman},\
  and\ \citenamefont {Porto}}]{Lin2009}%
  \BibitemOpen
  \bibfield  {author} {\bibinfo {author} {\bibfnamefont {Y.-J.}\ \bibnamefont
  {Lin}}, \bibinfo {author} {\bibfnamefont {A.~R.}\ \bibnamefont {Perry}},
  \bibinfo {author} {\bibfnamefont {R.~L.}\ \bibnamefont {Compton}}, \bibinfo
  {author} {\bibfnamefont {I.~B.}\ \bibnamefont {Spielman}}, \ and\ \bibinfo
  {author} {\bibfnamefont {J.~V.}\ \bibnamefont {Porto}},\ }\href {\doibase
  10.1103/PhysRevA.79.063631} {\bibfield  {journal} {\bibinfo  {journal}
  {Physical Review A (Atomic, Molecular, and Optical Physics)}\ }\textbf
  {\bibinfo {volume} {79}},\ \bibinfo {pages} {063631} (\bibinfo {year}
  {2009})}\BibitemShut {NoStop}%
\bibitem [{\citenamefont {Campbell}\ and\ \citenamefont
  {Spielman}(2016)}]{Campbell2016a}%
  \BibitemOpen
  \bibfield  {author} {\bibinfo {author} {\bibfnamefont {D.~L.}\ \bibnamefont
  {Campbell}}\ and\ \bibinfo {author} {\bibfnamefont {I.~B.}\ \bibnamefont
  {Spielman}},\ }\href {\doibase 10.1088/1367-2630/18/3/033035} {\bibfield
  {journal} {\bibinfo  {journal} {New Journal of Physics}\ }\textbf {\bibinfo
  {volume} {18}},\ \bibinfo {pages} {033035} (\bibinfo {year}
  {2016})}\BibitemShut {NoStop}%
\bibitem [{\citenamefont {Trypogeorgos}\ \emph {et~al.}(2018)\citenamefont
  {Trypogeorgos}, \citenamefont {Vald\'es-Curiel}, \citenamefont {Lundblad},\
  and\ \citenamefont {Spielman}}]{Trypogeorgos2018}%
  \BibitemOpen
  \bibfield  {author} {\bibinfo {author} {\bibfnamefont {D.}~\bibnamefont
  {Trypogeorgos}}, \bibinfo {author} {\bibfnamefont {A.}~\bibnamefont
  {Vald\'es-Curiel}}, \bibinfo {author} {\bibfnamefont {N.}~\bibnamefont
  {Lundblad}}, \ and\ \bibinfo {author} {\bibfnamefont {I.~B.}\ \bibnamefont
  {Spielman}},\ }\href {\doibase 10.1103/PhysRevA.97.013407} {\bibfield
  {journal} {\bibinfo  {journal} {Phys. Rev. A}\ }\textbf {\bibinfo {volume}
  {97}},\ \bibinfo {pages} {013407} (\bibinfo {year} {2018})}\BibitemShut
  {NoStop}%
\bibitem [{\citenamefont {Anderson}\ \emph {et~al.}(2018)\citenamefont
  {Anderson}, \citenamefont {Kewming},\ and\ \citenamefont
  {Turner}}]{Anderson2018}%
  \BibitemOpen
  \bibfield  {author} {\bibinfo {author} {\bibfnamefont {R.~P.}\ \bibnamefont
  {Anderson}}, \bibinfo {author} {\bibfnamefont {M.~J.}\ \bibnamefont
  {Kewming}}, \ and\ \bibinfo {author} {\bibfnamefont {L.~D.}\ \bibnamefont
  {Turner}},\ }\href {\doibase 10.1103/PhysRevA.97.013408} {\bibfield
  {journal} {\bibinfo  {journal} {Phys. Rev. A}\ }\textbf {\bibinfo {volume}
  {97}},\ \bibinfo {pages} {013408} (\bibinfo {year} {2018})}\BibitemShut
  {NoStop}%
\bibitem [{\citenamefont {Lamporesi}\ \emph {et~al.}(2010)\citenamefont
  {Lamporesi}, \citenamefont {Catani}, \citenamefont {Barontini}, \citenamefont
  {Nishida}, \citenamefont {Inguscio},\ and\ \citenamefont
  {Minardi}}]{lamporesi_scattering_2010}%
  \BibitemOpen
  \bibfield  {author} {\bibinfo {author} {\bibfnamefont {G.}~\bibnamefont
  {Lamporesi}}, \bibinfo {author} {\bibfnamefont {J.}~\bibnamefont {Catani}},
  \bibinfo {author} {\bibfnamefont {G.}~\bibnamefont {Barontini}}, \bibinfo
  {author} {\bibfnamefont {Y.}~\bibnamefont {Nishida}}, \bibinfo {author}
  {\bibfnamefont {M.}~\bibnamefont {Inguscio}}, \ and\ \bibinfo {author}
  {\bibfnamefont {F.}~\bibnamefont {Minardi}},\ }\href {\doibase
  10.1103/PhysRevLett.104.153202} {\bibfield  {journal} {\bibinfo  {journal}
  {Phys. Rev. Lett.}\ }\textbf {\bibinfo {volume} {104}},\ \bibinfo {pages}
  {153202} (\bibinfo {year} {2010})}\BibitemShut {NoStop}%
\bibitem [{\citenamefont {Leonard}\ \emph {et~al.}(2015)\citenamefont
  {Leonard}, \citenamefont {Fallon}, \citenamefont {Sackett},\ and\
  \citenamefont {Safronova}}]{leonard_high-precision_2015}%
  \BibitemOpen
  \bibfield  {author} {\bibinfo {author} {\bibfnamefont {R.~H.}\ \bibnamefont
  {Leonard}}, \bibinfo {author} {\bibfnamefont {A.~J.}\ \bibnamefont {Fallon}},
  \bibinfo {author} {\bibfnamefont {C.~A.}\ \bibnamefont {Sackett}}, \ and\
  \bibinfo {author} {\bibfnamefont {M.~S.}\ \bibnamefont {Safronova}},\ }\href
  {\doibase 10.1103/PhysRevA.92.052501} {\bibfield  {journal} {\bibinfo
  {journal} {Phys. Rev. A}\ }\textbf {\bibinfo {volume} {92}},\ \bibinfo
  {pages} {052501} (\bibinfo {year} {2015})}\BibitemShut {NoStop}%
\bibitem [{Note1()}]{Note1}%
  \BibitemOpen
  \bibinfo {note} {The added $\omega _Z$ in this expression as compared to
  Fig.~\ref {fig:setup}a is removed by a rotating frame transformation present
  in our RF dressed synthetic clock states.}\BibitemShut {Stop}%
\bibitem [{\citenamefont {Herold}\ \emph {et~al.}(2012)\citenamefont {Herold},
  \citenamefont {Vaidya}, \citenamefont {Li}, \citenamefont {Rolston},
  \citenamefont {Porto},\ and\ \citenamefont {Safronova}}]{Herold2012}%
  \BibitemOpen
  \bibfield  {author} {\bibinfo {author} {\bibfnamefont {C.~D.}\ \bibnamefont
  {Herold}}, \bibinfo {author} {\bibfnamefont {V.~D.}\ \bibnamefont {Vaidya}},
  \bibinfo {author} {\bibfnamefont {X.}~\bibnamefont {Li}}, \bibinfo {author}
  {\bibfnamefont {S.~L.}\ \bibnamefont {Rolston}}, \bibinfo {author}
  {\bibfnamefont {J.~V.}\ \bibnamefont {Porto}}, \ and\ \bibinfo {author}
  {\bibfnamefont {M.~S.}\ \bibnamefont {Safronova}},\ }\href {\doibase
  10.1103/PhysRevLett.109.243003} {\bibfield  {journal} {\bibinfo  {journal}
  {Phys. Rev. Lett.}\ }\textbf {\bibinfo {volume} {109}},\ \bibinfo {pages}
  {243003} (\bibinfo {year} {2012})}\BibitemShut {NoStop}%
\bibitem [{\citenamefont {Browaeys}\ \emph {et~al.}(2005)\citenamefont
  {Browaeys}, \citenamefont {H{\"a}ffner}, \citenamefont {McKenzie},
  \citenamefont {Rolston}, \citenamefont {Helmerson},\ and\ \citenamefont
  {Phillips}}]{browaeys_transport_2005}%
  \BibitemOpen
  \bibfield  {author} {\bibinfo {author} {\bibfnamefont {A.}~\bibnamefont
  {Browaeys}}, \bibinfo {author} {\bibfnamefont {H.}~\bibnamefont
  {H{\"a}ffner}}, \bibinfo {author} {\bibfnamefont {C.}~\bibnamefont
  {McKenzie}}, \bibinfo {author} {\bibfnamefont {S.~L.}\ \bibnamefont
  {Rolston}}, \bibinfo {author} {\bibfnamefont {K.}~\bibnamefont {Helmerson}},
  \ and\ \bibinfo {author} {\bibfnamefont {W.~D.}\ \bibnamefont {Phillips}},\
  }\href {\doibase 10.1103/PhysRevA.72.053605} {\bibfield  {journal} {\bibinfo
  {journal} {Physical Review A}\ }\textbf {\bibinfo {volume} {72}},\ \bibinfo
  {pages} {053605} (\bibinfo {year} {2005})}\BibitemShut {NoStop}%
\bibitem [{\citenamefont {Peik}\ \emph {et~al.}(1997)\citenamefont {Peik},
  \citenamefont {Ben~Dahan}, \citenamefont {Bouchoule}, \citenamefont
  {Castin},\ and\ \citenamefont {Salomon}}]{peik_bloch_1997}%
  \BibitemOpen
  \bibfield  {author} {\bibinfo {author} {\bibfnamefont {E.}~\bibnamefont
  {Peik}}, \bibinfo {author} {\bibfnamefont {M.}~\bibnamefont {Ben~Dahan}},
  \bibinfo {author} {\bibfnamefont {I.}~\bibnamefont {Bouchoule}}, \bibinfo
  {author} {\bibfnamefont {Y.}~\bibnamefont {Castin}}, \ and\ \bibinfo {author}
  {\bibfnamefont {C.}~\bibnamefont {Salomon}},\ }\href {\doibase
  10.1103/PhysRevA.55.2989} {\bibfield  {journal} {\bibinfo  {journal}
  {Physical Review A}\ }\textbf {\bibinfo {volume} {55}},\ \bibinfo {pages}
  {2989} (\bibinfo {year} {1997})}\BibitemShut {NoStop}%
\bibitem [{\citenamefont {Su}\ \emph {et~al.}(1979)\citenamefont {Su},
  \citenamefont {Schrieffer},\ and\ \citenamefont {Heeger}}]{Su1979}%
  \BibitemOpen
  \bibfield  {author} {\bibinfo {author} {\bibfnamefont {W.~P.}\ \bibnamefont
  {Su}}, \bibinfo {author} {\bibfnamefont {J.~R.}\ \bibnamefont {Schrieffer}},
  \ and\ \bibinfo {author} {\bibfnamefont {A.~J.}\ \bibnamefont {Heeger}},\
  }\href {\doibase 10.1103/PhysRevLett.42.1698} {\bibfield  {journal} {\bibinfo
   {journal} {Phys. Rev. Lett.}\ }\textbf {\bibinfo {volume} {42}},\ \bibinfo
  {pages} {1698} (\bibinfo {year} {1979})}\BibitemShut {NoStop}%
\bibitem [{\citenamefont {Vald{\'e}s-Curiel}\ \emph {et~al.}(2017)\citenamefont
  {Vald{\'e}s-Curiel}, \citenamefont {Trypogeorgos}, \citenamefont {Marshall},\
  and\ \citenamefont {Spielman}}]{Valdes-Curiel2017}%
  \BibitemOpen
  \bibfield  {author} {\bibinfo {author} {\bibfnamefont {A.}~\bibnamefont
  {Vald{\'e}s-Curiel}}, \bibinfo {author} {\bibfnamefont {D.}~\bibnamefont
  {Trypogeorgos}}, \bibinfo {author} {\bibfnamefont {E.~E.}\ \bibnamefont
  {Marshall}}, \ and\ \bibinfo {author} {\bibfnamefont {I.~B.}\ \bibnamefont
  {Spielman}},\ }\href {\doibase 10.1088/1367-2630/aa6279} {\bibfield
  {journal} {\bibinfo  {journal} {New Journal of Physics}\ }\textbf {\bibinfo
  {volume} {19}},\ \bibinfo {pages} {033025} (\bibinfo {year}
  {2017})}\BibitemShut {NoStop}%
\bibitem [{\citenamefont {Cooper}\ and\ \citenamefont
  {Moessner}(2012)}]{Cooper2012}%
  \BibitemOpen
  \bibfield  {author} {\bibinfo {author} {\bibfnamefont {N.~R.}\ \bibnamefont
  {Cooper}}\ and\ \bibinfo {author} {\bibfnamefont {R.}~\bibnamefont
  {Moessner}},\ }\href@noop {} {\bibfield  {journal} {\bibinfo  {journal}
  {Physical Review Letters}\ }\textbf {\bibinfo {volume} {109}},\ \bibinfo
  {pages} {265301} (\bibinfo {year} {2012})}\BibitemShut {NoStop}%
\bibitem [{\citenamefont {Lu}\ \emph {et~al.}(2018)\citenamefont {Lu},
  \citenamefont {Gao},\ and\ \citenamefont {Wang}}]{Lu2018}%
  \BibitemOpen
  \bibfield  {author} {\bibinfo {author} {\bibfnamefont {L.}~\bibnamefont
  {Lu}}, \bibinfo {author} {\bibfnamefont {H.}~\bibnamefont {Gao}}, \ and\
  \bibinfo {author} {\bibfnamefont {Z.}~\bibnamefont {Wang}},\ }\href {\doibase
  10.1038/s41467-018-07817-3} {\bibfield  {journal} {\bibinfo  {journal}
  {Nature Communications}\ }\textbf {\bibinfo {volume} {9}},\ \bibinfo {pages}
  {5384} (\bibinfo {year} {2018})}\BibitemShut {NoStop}%
\bibitem [{\citenamefont {Sugawa}\ \emph {et~al.}(2018)\citenamefont {Sugawa},
  \citenamefont {Salces-Carcoba}, \citenamefont {Perry}, \citenamefont {Yue},\
  and\ \citenamefont {Spielman}}]{Sugawa2018}%
  \BibitemOpen
  \bibfield  {author} {\bibinfo {author} {\bibfnamefont {S.}~\bibnamefont
  {Sugawa}}, \bibinfo {author} {\bibfnamefont {F.}~\bibnamefont
  {Salces-Carcoba}}, \bibinfo {author} {\bibfnamefont {A.~R.}\ \bibnamefont
  {Perry}}, \bibinfo {author} {\bibfnamefont {Y.}~\bibnamefont {Yue}}, \ and\
  \bibinfo {author} {\bibfnamefont {I.~B.}\ \bibnamefont {Spielman}},\ }\href
  {\doibase 10.1126/science.aam9031} {\bibfield  {journal} {\bibinfo  {journal}
  {Science}\ }\textbf {\bibinfo {volume} {360}},\ \bibinfo {pages} {1429}
  (\bibinfo {year} {2018})},\ \Eprint
  {http://arxiv.org/abs/http://science.sciencemag.org/content/360/6396/1429.full.pdf}
  {http://science.sciencemag.org/content/360/6396/1429.full.pdf} \BibitemShut
  {NoStop}%
\bibitem [{\citenamefont {Kennedy}\ \emph {et~al.}(2013)\citenamefont
  {Kennedy}, \citenamefont {Siviloglou}, \citenamefont {Miyake}, \citenamefont
  {Burton},\ and\ \citenamefont {Ketterle}}]{Kennedy2013}%
  \BibitemOpen
  \bibfield  {author} {\bibinfo {author} {\bibfnamefont {C.~J.}\ \bibnamefont
  {Kennedy}}, \bibinfo {author} {\bibfnamefont {G.~A.}\ \bibnamefont
  {Siviloglou}}, \bibinfo {author} {\bibfnamefont {H.}~\bibnamefont {Miyake}},
  \bibinfo {author} {\bibfnamefont {W.~C.}\ \bibnamefont {Burton}}, \ and\
  \bibinfo {author} {\bibfnamefont {W.}~\bibnamefont {Ketterle}},\ }\href
  {\doibase 10.1103/PhysRevLett.111.225301} {\bibfield  {journal} {\bibinfo
  {journal} {Phys. Rev. Lett.}\ }\textbf {\bibinfo {volume} {111}},\ \bibinfo
  {pages} {225301} (\bibinfo {year} {2013})}\BibitemShut {NoStop}%
\end{thebibliography}%

\end{document}